\documentclass[twocolumn]{aastex631}
\usepackage{xspace}
\usepackage{soul}
\usepackage{mathtools}
\newcommand{\swift}{{\em Swift}\xspace}

\newcommand{\tninty}{{$T_{\rm 90}$}\xspace}
\newcommand{\bat}{{\em Swift}/BAT\xspace}
\newcommand{\mvts}{{$t_{\rm mvts}$}\xspace}

\newcommand{\sw}[1]{\texttt{#1}}
\newcommand\T{\rule{0pt}{4ex}}

\begin{document}

\title[GRB211211A]{
Evidence for two distinct populations of kilonova-associated Gamma Ray Bursts}
\author[0000-0001-9868-9042]{Dimple}
\affiliation{Aryabhatta Research Institute of Observational Sciences (ARIES), Manora Peak, Nainital-263002, India. }
\affiliation{Department of Physics, Deen Dayal Upadhyaya Gorakhpur University, Gorakhpur-273009, India.}
\author[0000-0003-1637-267X]{K. Misra}
\affiliation{Aryabhatta Research Institute of Observational Sciences (ARIES), Manora Peak, Nainital-263002, India. }
\author[0000-0002-6960-8538]{K. G. Arun}
\affiliation{Chennai Mathematical Institute, Siruseri, 603103 Tamilnadu, India.}
\affiliation{Institute for Gravitation and the Cosmos, Department of Physics, Penn State University, University Park, \\Pennsylvania 16802, USA}

\begin{abstract}
Identification of Gamma Ray Burst (GRB) progenitors based on the duration of their prompt emission ($T_{90}$) has faced several roadblocks recently. Long-duration GRBs (with $T_{90} > 2s$) have traditionally been thought to be originating from the collapse of massive stars, and the short-duration ones (with $T_{90} < 2s$) from compact binary mergers. However, recent observations of a long GRB associated with a kilonova (KN) and a short GRB with supernova (SN) association demand a more detailed classification of the GRB population. In this {\it Letter}, we focus on GRBs associated with KNe, believed to be originating from mergers of binaries involving neutron stars (NS). We make use of the GRB prompt emission light curves of \bat 2022 GRB catalog and employ machine learning algorithms to study the classification of GRB progenitors. Our analysis reveals that there are five distinct clusters of GRBs, of which the KN-associated GRBs are located in two separate clusters indicating they may have been produced by different progenitors. We argue that these clusters may be due to subclasses of binary neutron star (BNS) and/or neutron star--black hole (NS-BH) mergers. We also discuss the implications of these findings for future gravitational-wave (GW) observations and how those observations may help in understanding these clusters better.
\end{abstract}

\keywords{GRBs --- Kilonovae -- BNS mergers --- NS-BH mergers --- PCA --- Machine Learning --- tSNE --- UMAP}

\section{Introduction} \label{sec:intro}
GRBs are traditionally classified as long and short bursts based on the bimodal distribution of their duration and their distinct spectral hardness ratios \citep{Kouveliotou1993}. Long bursts, with \tninty $> 2 s$ (the duration over which 90\% of the energy in prompt emission is released), have been argued to originate from the collapse of massive stars \citep{Woosley1993,Paczyski1998,Mszros2006,Hjorth2012,Woosley2006}. This argument is supported by the detection of broad-lined type Ic SNe associated with several of the long GRBs \citep{Galama_1998,Kulkarni1998,Stanek_2003,Cano2017}. On the other hand, short GRBs have been argued to arise from compact object mergers \citep{Paczynski1986,Eichler_1989,Meszaros1992}. The coincident discovery of GRB~170817A and GW~170817 confirmed that BNS mergers do produce short GRBs \citep{Abbott2017,Goldstein2017}. Kilonovae (KNe), argued to be produced by {\it r-process} nucleosynthesis during compact object mergers (containing a NS), have been observed in association with at least three of the short GRBs - 130603B \citep{Tanvir2013,Berger2013}, 160821B \citep{Kasliwal2017,Lamb2019,Troja2019}, and 170817A \citep{Valenti_2017,Abbott2017b}, further strengthening the conjecture that mergers of compact binaries can produce short GRBs. 

However, there have been many instances where this canonical association of long GRBs to massive star collapse and short GRBs to compact binary mergers has been proven to be incorrect. For instance, GRB~200826A despite being short, had associated SN \citep{Rossi2022,Ahumada2021,zhang_2021} and long GRBs~060614 and 060505 \citep{Fynbo_2006}, had no accompanying SNe, and even more interestingly GRBs~060614 \citep{Yang2015} and 211227A \citep{L2022}, yielded evidence for accompanying KNe. In recent studies, collapsar signatures have been noticed in high-redshift short GRBs \citep{Dimple2022b}.

The recent discovery of a long burst, GRB~211211A, further complicates the classification problem demanding a detailed re-examination of the paradigm. A confirmed KN counterpart \citep{Rastinejad2022} was identified with this burst despite being a long GRB.  The host galaxy properties were found to be similar to short GRB hosts rather than long GRB hosts \citep{Troja2022,Yang2022}. The observed large offset from the center of the host and lack of spatial association with any star-forming region further strengthens the claim of a compact binary merger as a progenitor to this burst \citep{Fruchter2006,Berger2013}. Additionally, there is evidence of GeV emission observed in the prompt phase of the burst. \cite{2022Natur.612..236M} suggested that this GeV emission could be caused by the inverse Compton effect, where the jet interacts with an external source of photons. They proposed that the KN emission observed could provide the necessary seed photons for the GeV emission.

Various progenitor models have been put forward to explain the origin of GRB~211211A. \cite{Yang2022} proposed a neutron star-white dwarf (NS-WD) merger with a rapidly spinning magnetar as the central engine. However, \cite{Gompertz2022} suggest that the fast evolution of the burst light curves cannot be explained by a NS-WD merger (or a SN), though their study supports a merger origin of the burst. The presence of Quasi-periodic oscillations (QPOs) in the prompt emission light curve of the burst has been argued to favor a magnetar progenitor \citep{Xiao_2022}. \cite{Troja2022} claim that the time scales and colors of the magnetar-induced KN do not match the observed light curves, proposing it to be a hybrid event (that shares similarity to both long and short GRBs) originating from a compact object merger. It was revealed in the studies by \cite{Barnes2023} and \cite{Kunert2023} that a collapsar origin is unlikely, and the origin of the burst is consistent with a BNS or NS-BH merger, with a mild preference for a BNS merger. On the other hand, \cite{Zhu2022} reported the fallback signatures in GRB~211211A, suggesting a NS-BH merger origin. Recent studies by \cite{Zhong_2023} claim that GRB~211211A could arise from a NS-WD merger if it produces a magnetar as remnant. They argue that the optical afterglows, along with KN-like emission, can be well-modeled by combining the standard forward shock with the radioactive decay power of $\rm \prescript{56}{}{Ni}$ adding a rotational power input from the post-merger magnetar. 

All these studies have strongly favored a merger origin for the long GRB~211211A in sharp contradiction to it being of collapsar origin as is canonically thought for long GRBs. Previously, two other long GRBs~060614 \citep{Yang2015} and 211227A \citep{L2022} have been asserted to show signatures of merger origin. It is, therefore, important to ask whether all GRBs with a possible KNe-association share similar features and have a common progenitor. 

Arguably, the best approach to answer this question is to study a population of GRBs detected by an instrument and analyze the corresponding prompt emission light curves of the GRBs. The combined use of prompt emission light curves and state-of-the-art machine learning techniques could help in identifying any global structures within the GRB population and thereby give us some vital clues on distinct progenitor classes of GRBs. However, analyzing a large sample of GRB light curves can significantly increase the dimensionality of the data, making analysis difficult. Traditional dimension reduction methods, such as Principal Component Analysis (PCA), can address the high dimensionality issue, but it has limitations when interpreting the Principal Components (PCs) and using them for clustering. To overcome these limitations, non-linear methods like t-distributed Stochastic Neighbor Embedding (tSNE; \citealt{JMLR:v9:vandermaaten08a, JMLR:v15:vandermaaten14a}) and Uniform Manifold Approximation and Projection (UMAP; \citealt{McInnes2018}) can be employed. 

\cite{Christian2020} and \cite{Steinhardt2023} have used tSNE and UMAP, respectively, to identify clustering in the population of GRBs. \cite{Luo_2022} also made use of supervised machine learning techniques to distinguish between different progenitor systems of GRBs. However, these studies focused more on the general classification of GRBs based on their light curves and did not survey any specific subpopulation, such as that of the KN-associated GRBs.  
Recently, \cite{Keneth_2023} has used tSNE to investigate the extended emission GRBs. However, all these approaches used tSNE or UMAP alone, which preserves the global structures but not the local structures in the data. Instead, \citet{Kobak2021} have shown that initializing tSNE/UMAP with PCA can improve its ability to capture both the global and local structures in the data. 

Motivated by the recent advances in the development of machine learning algorithms and the puzzles posed by GRB~211211A and similar GRBs, in this {\it Letter}, we analyze the landscape of KN-associated GRBs in the \bat catalog\footnote{\url{https://swift.gsfc.nasa.gov/results/batgrbcat/}} \citep{Lien_2016} using UMAP and tSNE with a PCA initialization. Unlike the previous works, our goal is to identify subclusters in the GRB population. Specifically, we are interested in identifying subclasses of KN-associated GRBs and understanding their possible progenitors.

\section{Methodology }
\label{methods}
\subsection{Use of Machine Learning tools}
Unsupervised machine learning techniques are widely used for clustering and dimensionality reduction in diverse areas of science. Such methods have the ability to reveal hidden patterns in the data associated with complex phenomena. In our context, the underlying physics associated with the prompt emission of GRBs is extremely complex and not well understood. It is possible to use some of the well-known GRBs, whose properties are studied extensively, to assist the interpretation of the classification derived from these algorithms. This is the approach we take in this work where the techniques of tSNE and UMAP with PCA initialization are used to cluster a set of GRB prompt emission light curves, and then the known properties of a subset of the GRBs are incorporated to provide a basis for the interpretation. 

tSNE is a non-linear dimensionality reduction technique that helps in visualizing the data using stochastic embedding of neighborhoods \citep{JMLR:v9:vandermaaten08a, JMLR:v15:vandermaaten14a}. It models the probability distributions of pairwise similarities in high and lower-dimensional spaces and minimizes their differences using Kullback-Leibler divergence \citep{kullback1951}. Initially, the algorithm randomly positions points in the high-dimensional space, then iteratively moves each point towards a position in the lower-dimensional space that best aligns with its neighboring points. This movement is guided by a divergence function that is minimized when the distribution of points in the lower-dimensional space matches that in the high-dimensional space. The hyperparameters of the algorithm control the visualization of the data, particularly the \sw{perplexity} parameter, which determines the number of nearest neighbors to consider when calculating similarity distribution. A higher \sw{perplexity} value takes more neighbors into account, producing a smoother-looking embedding with greater emphasis on global structure. Conversely, a lower \sw{perplexity} value produces an embedding with a more detailed local structure but less emphasis on global structure. Choosing the right \sw{perplexity} value is crucial for capturing both local and global structure. For a given problem, one tune this parameter to get the balance between local and global structures that is necessary for the visualization of the data.

Uniform Manifold Approximation and Projection, on the other hand, uses more abstract mathematical concepts from topology for dimension reduction, and therefore the method is more deterministic than tSNE, especially when it comes to defining notions of distance  \citep{McInnes2018}. The algorithm aims to reduce the dimensionality of data while preserving its local structure.
The algorithm consists of two main steps: first, constructing a high-dimensional representation of the data by creating a similarity matrix between all points in the data set. Second, reducing the dimensionality of the data while retaining its local structure by finding a low-dimensional projection that preserves the distances between all points in the similarity matrix. The algorithm has several parameters, but the number of nearest neighbors (\sw{n\_neighbors}, similar to \sw{perplexity}) and minimum distance (\sw{min\_dist}, how closely the method packs points which are close to each other) are particularly important in balancing local and global structure in the final projection. Increasing \sw{n\_neighbors} connects more neighboring points, resulting in a more precise representation of the overall structure of data, but may not capture the local structures in the data and therefore increase computational complexity. On the other hand, increasing \sw{min\_dist} results in a more spread-out representation of the data, reducing data clumping but potentially leading to a loss of global structure. By construction, UMAP has a better ability to preserve global distances compared to tSNE. Further, it is found that UMAP is more time-efficient than t-SNE with regard to computation time. 

In this work, we analyze the landscape of \swift GRBs utilizing the prompt emission light curves in four energy bins with machine learning technique; PCA-tSNE/UMAP. Similarities between the light curves are used by these algorithms to produce two-dimensional maps of the GRBs, which are further clustered using \sw{AutoGMM} (a clustering algorithm to optimize the number of clusters in a given data). Then, we superimpose the known KNe/SNe cases on top of the identified clusters and infer interesting correlations and discuss their implications for multi-messenger astronomy involving GWs. Next, we turn our attention to the technical aspects of the analysis.
\subsection{Data standardization}
\label{data}
The \bat catalog 2022 consists of prompt emission light curves of 1525 GRBs detected between 17 December 2004 and 15 July 2022. The diverse data from \bat reveals varying durations, fluence, and start times of the bursts, which can lead to distraction in the algorithm. Hence, it is necessary to homogenize the data format before employing any machine learning program. To do this, we followed the standardizing procedure described in \cite{Christian2020}. The main steps involved are i) normalizing the light curves with the fluence, ii) shifting the light curves to the same start time, iii) padding the light curves with zero so that they have a similar length, and iv) performing the Discrete-time Fourier Transform (DTFT) to preserve the time delay information between the different light curves. The duration and fluence (corresponding to the best-fit model) values were adopted from the summary files provided in the \bat catalog. For some GRBs, the \tninty or fluence information was missing. Due to this, the standardization procedure could be carried out only for 1450 GRBs. We saved this standardized data in the form of a matrix with dimensions [1450 x 25883], where each row represents a GRB and contains corresponding light curves in four energy channels with a binning of 64 ms.
\begin{figure}[!h]
\centering
    \includegraphics[width=\columnwidth]{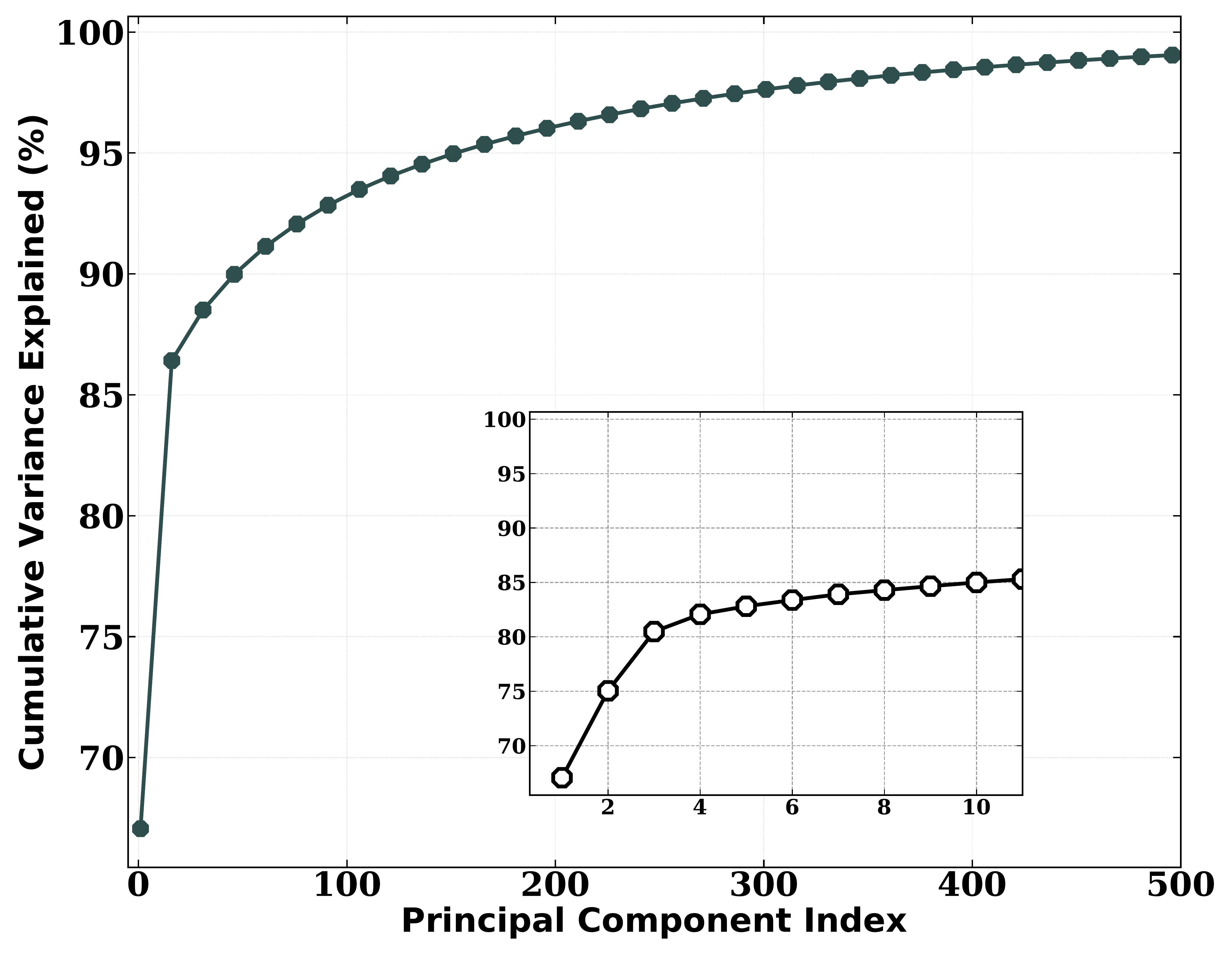}
    \caption{Scree plot presenting the contribution of PCs to explain the variance of the data. The leading four components explain  $\sim 82$\% of the variance (inset). The first 500 components explain 99\% of the variance.}
    \label{fig:scree-plot}
\end{figure}

\subsection{Dimensionality reduction}
\label{clustering}
Following the standardization of data, the subsequent step involves dimensionality reduction. To accomplish this, we first performed the PCA and generated the corresponding scree plot, which is a graphical representation explaining the variance of each PC. Fig.~\ref{fig:scree-plot} illustrates the scree plot, it can be observed that around 4 leading PCs account for approximately 80\% of the variance in the data. 
However, in order to avoid biases due to the incompleteness of the data,  we utilized the leading 500 PCs, which preserves approximately 99\% of the variance. This resulted in a matrix with dimensions of [1450 x 500], which we used to generate two-dimensional embeddings using tSNE and UMAP algorithms. 

\begin{figure*}
    \centering
    \includegraphics[width=\columnwidth]{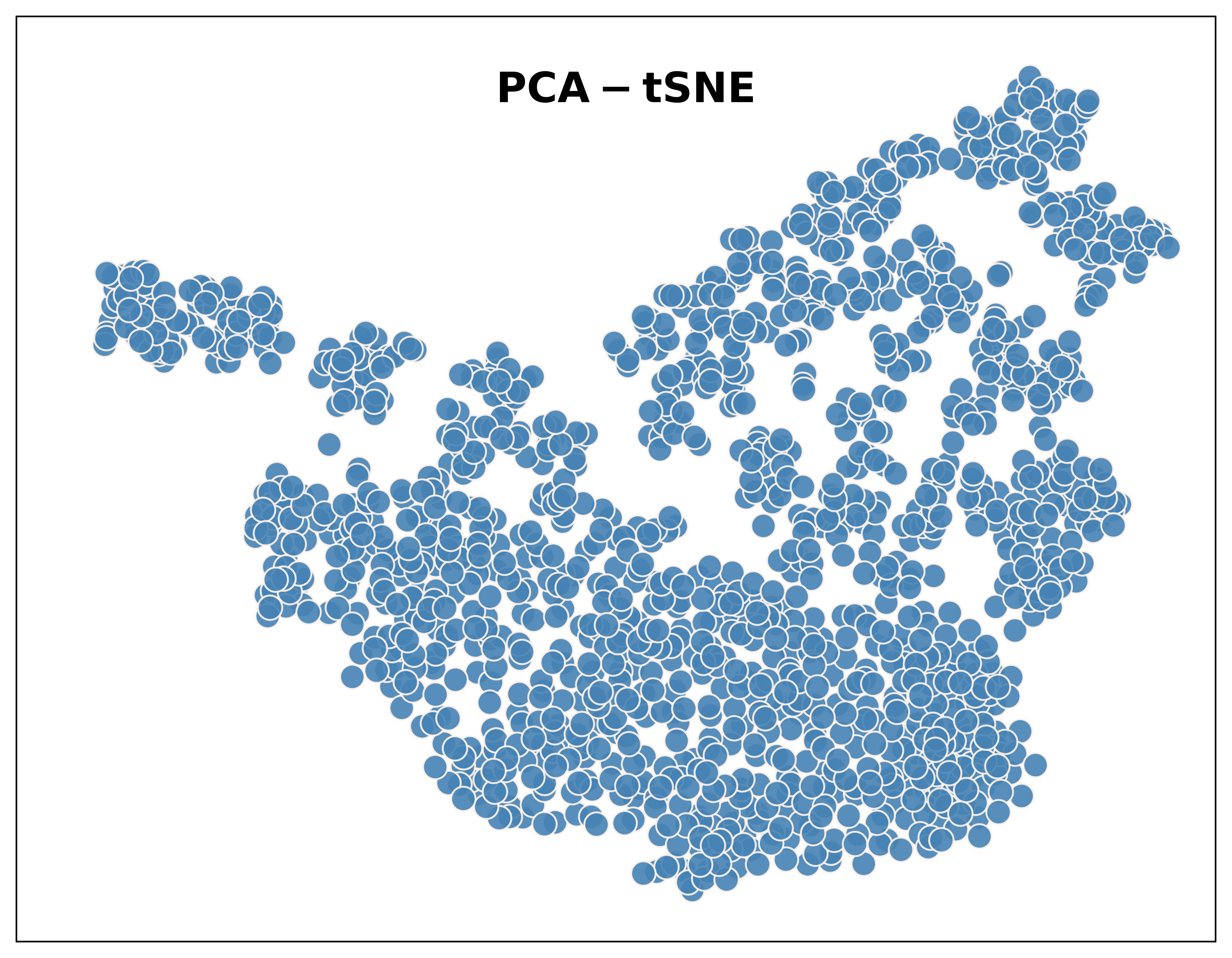}
    \includegraphics[width=\columnwidth]{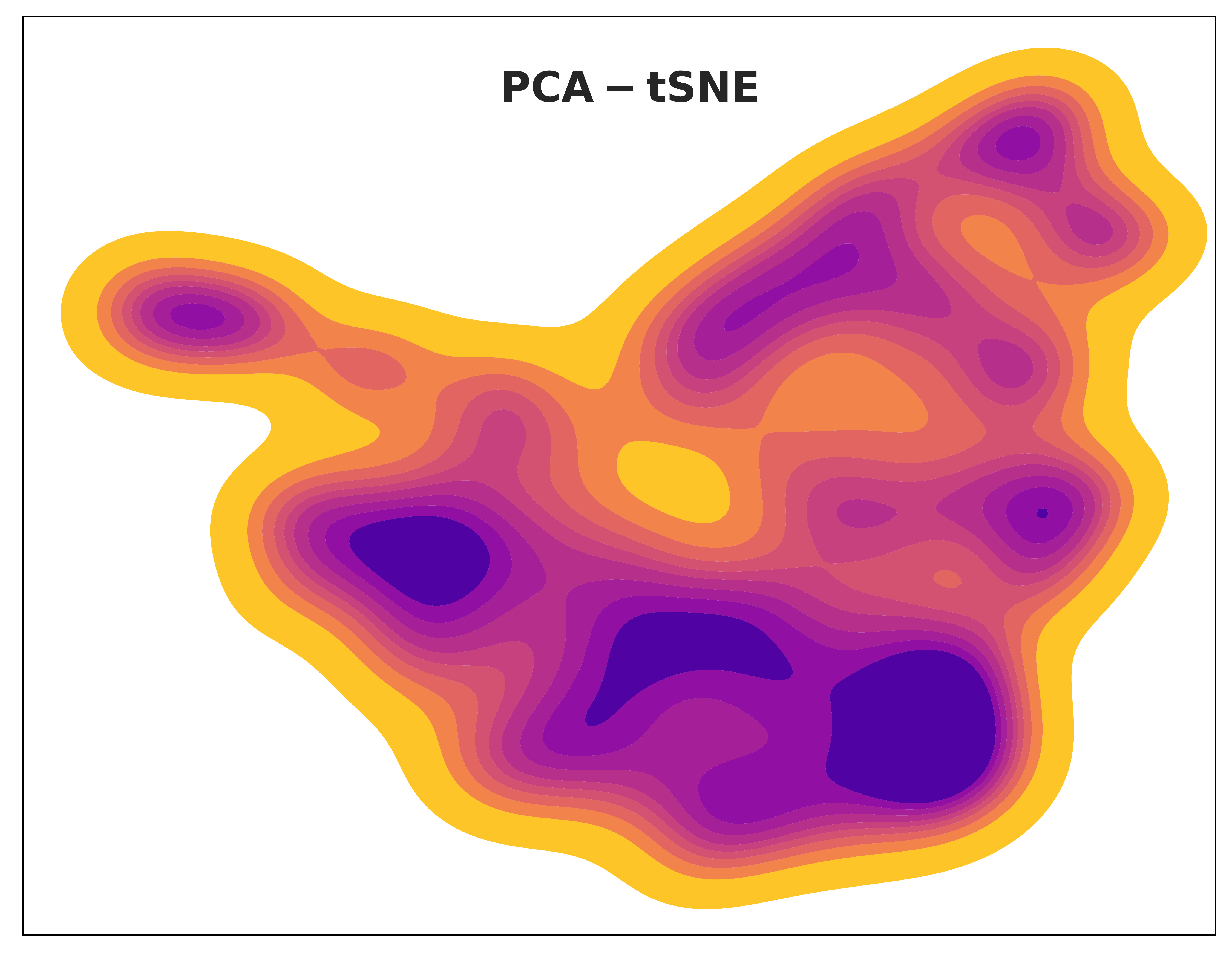}
    \includegraphics[width=\columnwidth]{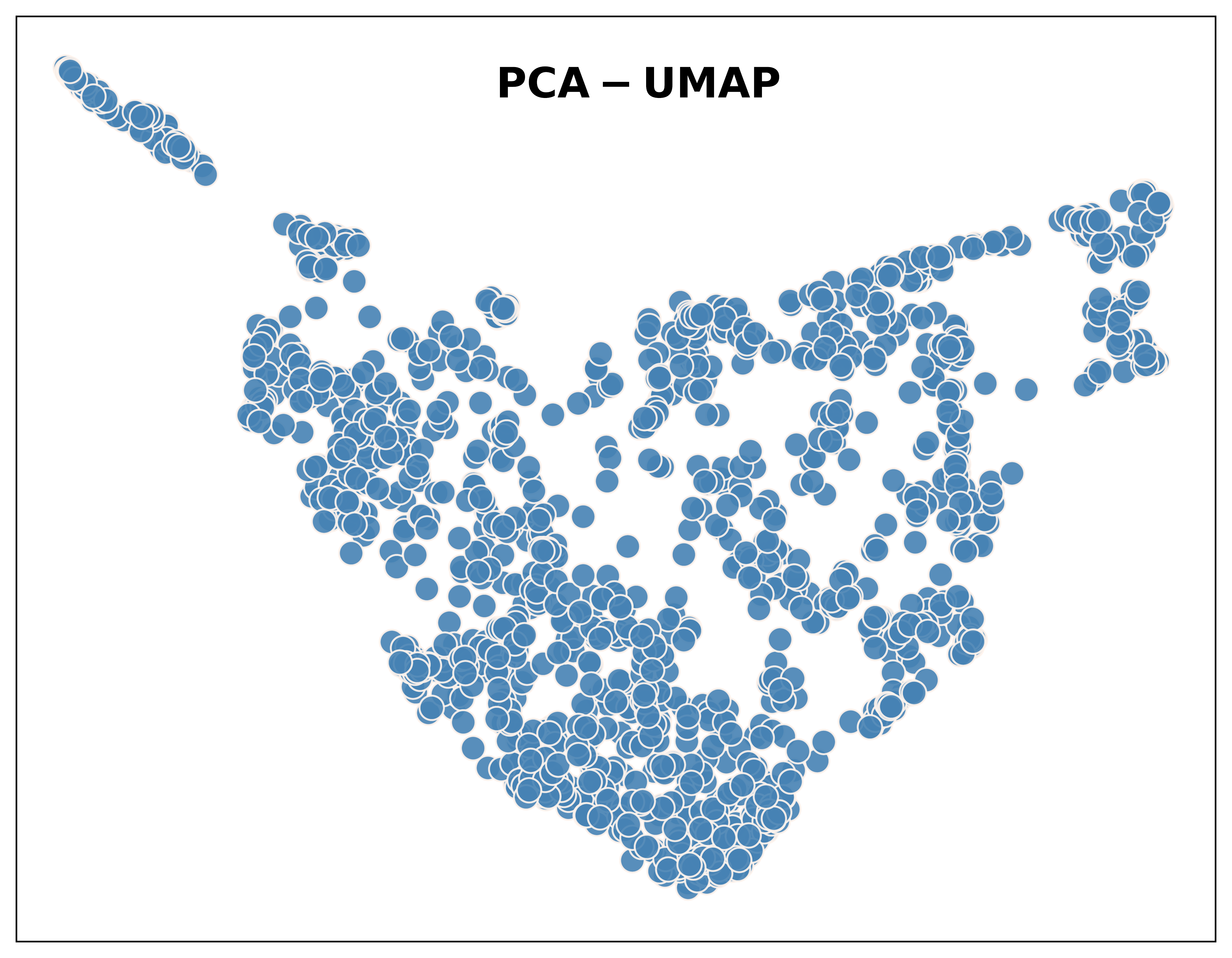}
    \includegraphics[width=\columnwidth]{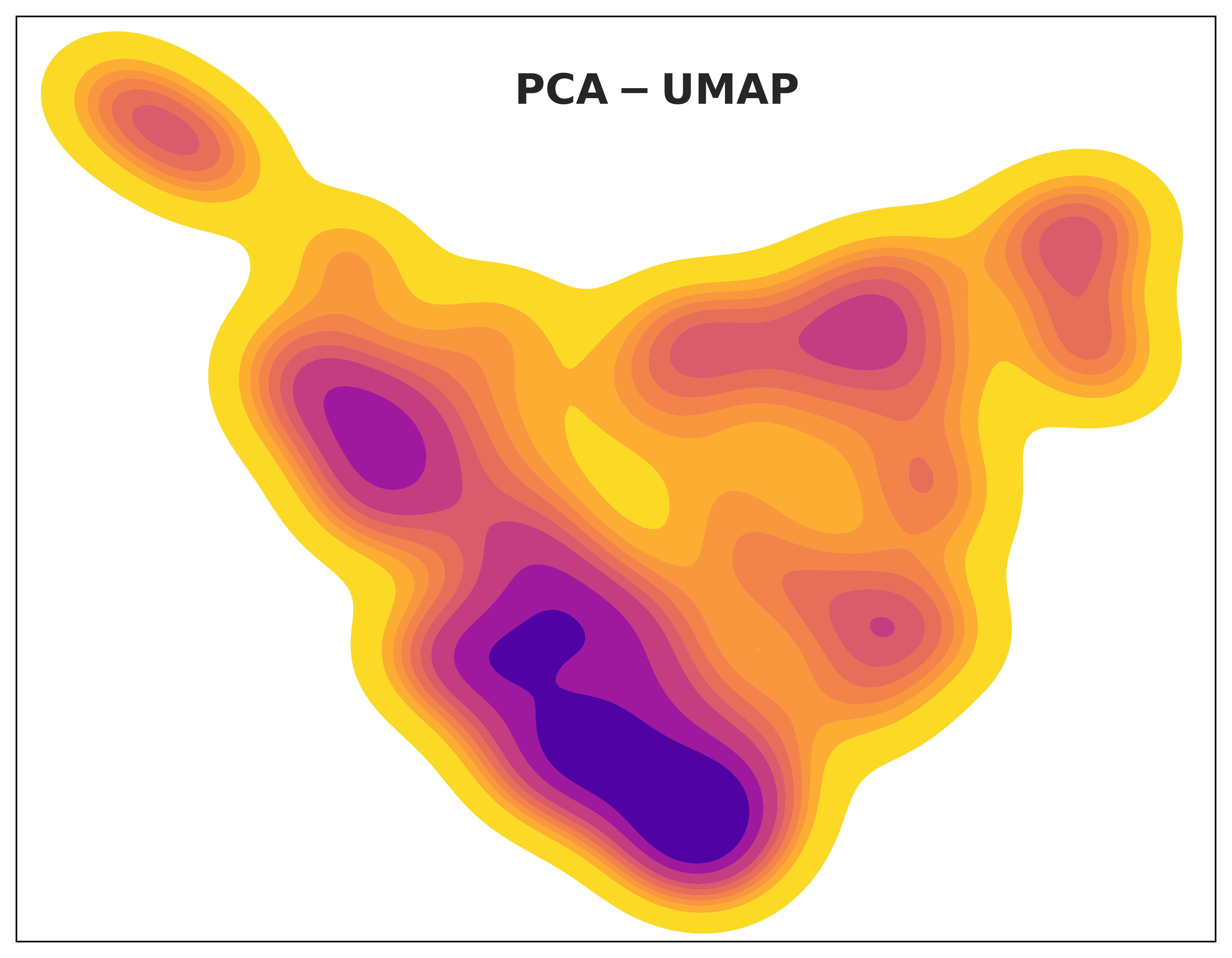}
    \caption{{\bf Top:} PCA-tSNE results depicting the two-dimensional embedding (left) and the respective density map (right, bluer regions correspond to a higher density of the bursts clustered at different locations on the map). {\bf Bottom:} Results for PCA-UMAP depicting embedding and density map.}
    \label{fig:PCA-tSNE/UMAP}
\end{figure*}

As previously stated, the tSNE and UMAP algorithms are sensitive to specific hyperparameters, and the proper selection of these key parameters is essential for obtaining a meaningful representation of the data in a two-dimensional space. We tuned these parameters after several iterations, we gradually increased the \sw{perplexity} from 5 to 50 (in increments of 5) in the case of tSNE and chose a value of 25, for which we get a clear separation of various clusters.  Fig. \ref{fig:PCA-tSNE/UMAP} shows the embeddings obtained using PCA-tSNE (top left) and the respective density plot (top right). 

For UMAP, we iteratively increased the \sw{n\_neighbor} from 5 to 50 (in increments of 5) and decreased \sw{min\_dist} from 0.1 to 0.005. We chose the parameters where we get a clear separation of various clusters. We fixed the \sw{n\_neighbors} to 25 and \sw{min\_dist} to 0.01. The 2-dimensional embedding (bottom left), along with the density plots (bottom right) obtained using these parameters, is shown in Fig. \ref{fig:PCA-tSNE/UMAP}. 
It should be noted that various choices of hyperparameters simply translate into  different projections of the data from higher dimensions to two dimensions. Those projections that are more useful for visualization of the data, depending on the context, are employed in the analysis.

\subsection{Clustering with AutoGMM}
\label{AutoGMM}
\begin{figure*}
\centering
    \includegraphics[scale=0.20]{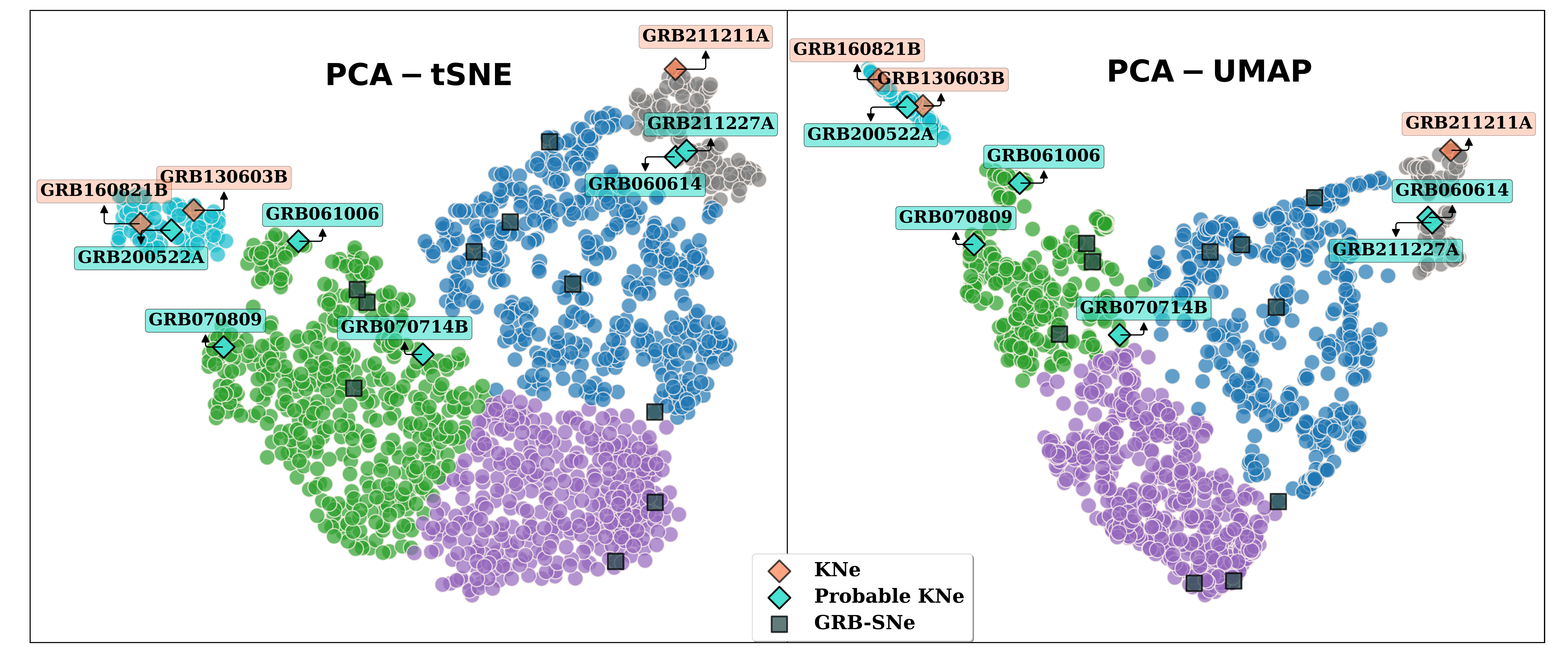}
    \caption{{\bf {Left:}} The locations of KN-associated GRBs on a two-dimensional embedding obtained using PCA-tSNE. The coral-colored labels represent the GRBs with confirmed KN association. The turquoise-colored labels are the GRBs with probable KN candidates. The KN-associated GRBs with long and short duration occupy two locations on the embedding (right and left corners). The brown-colored symbols show the location of GRB-SNe on the embedding. {\bf {Right:}} Same using PCA-UMAP. The embeddings are color-coded with respect to the five clusters identified by \sw{AutoGMM}.}
    \label{fig:kilonovae}
\end{figure*}

For a more flexible and accurate representation of the data and to identify the various clusters within the map, we used the \sw{AutoGMM} module available in Python \citep{Athey2019}. \sw{AutoGMM} is an automated approach to determine the optimal number of clusters in the data using GMM that postulates that data points originate from a combination of several Gaussian distributions. The algorithm begins by running the GMM clustering process for a range of different cluster numbers. The algorithm then tries to estimate the parameters of the Gaussian distributions that best fit the data, including the number of clusters in the dataset. The application of \sw{AutoGMM} to the two-dimensional embeddings obtained from PCA-tSNE and PCA-UMAP resulted in the identification of five clusters in the embedding (see Fig.~\ref{fig:kilonovae}).

Since both the algorithms PCA-tSNE and PCA-UMAP work independently and are governed by several hyperparameters, it is crucial to compare the results obtained from both methods. Therefore, we assess the similarity between the embeddings generated from these algorithms using \sw{flameplot}, as outlined in Appendix \ref{comparison_flameplot}.

\section{Results and Discussion}
\label{discussion}
\begin{figure*}
\centering
    \includegraphics[scale=0.36]{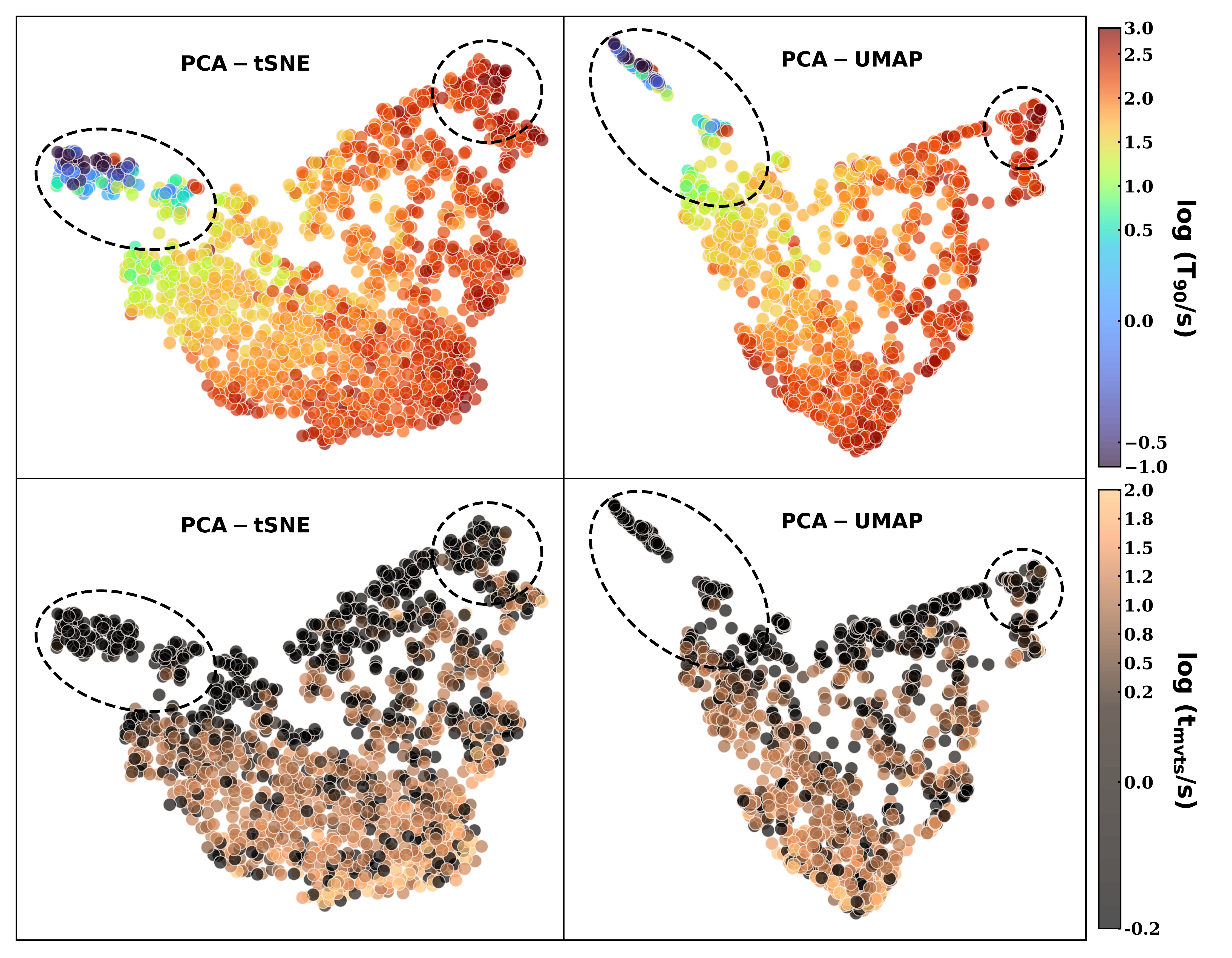}
    \caption{PCA-tSNE and PCA-UMAP embedding color-coded with \tninty (top) and \mvts (bottom). The encircled regions indicate the locations of KN-associated GRBs in the embeddings. 
    }
    \label{fig:comparison}
\end{figure*}

Fig.~\ref{fig:kilonovae} displays the two-dimensional embeddings obtained using PCA-tSNE and PCA-UMAP, color-coded with the clusters identified by \sw{AutoGMM}. It is conspicuous from the figure that the algorithm identifies five distinct clusters in both embeddings, indicating five different classes of GRBs in the \bat catalog 2022. This suggests that the results obtained from both the algorithms (PCA-tSNE and PCA-UMAP) are similar, which is also evident from their similarity score (see Fig.~\ref{flameplot}). Here, we will focus only on those GRBs with KN counterparts, and a detailed analysis of the full landscape will be reported elsewhere (Dimple et al., in preparation). 

We locate the confirmed and probable candidates of the KN-associated GRBs on the PCA-tSNE/UMAP embeddings. The sample of confirmed KN-associated cases include GRBs~130603B \citep{Berger2013,Tanvir2013}, 160821B \citep{Kasliwal2017,Lamb2019,Troja2019} and 211211A \citep{Rastinejad2022,Troja2022}. Although GRB~170817A has a confirmed KN-association, we do not have it in our sample as it was not detected by \bat. GRBs~060614 \citep{Yang2015}, 061006 \citep{Gao2017}, 070714B \citep{Gao2017}, 070809 \citep{Jin2019}, 200522A \citep{Fong2021,Oconnor_2021} and 211227A \citep{L2022} were discussed to be probable KN candidates and constitute the sample in this work. 

The KN-associated GRBs are seen to occupy two distinct regions on both embeddings (see Fig.~\ref{fig:kilonovae}). Six GRBs lie on the left side of the embedding (left cluster henceforth), and three others are on the right side (right cluster henceforth). Interestingly, the GRBs in the left cluster have been traditionally classified as short and those lying in the right cluster as  long based on the \tninty duration (see Fig.~\ref{fig:comparison}). Further, to better understand the landscape of GRBs, we track all the GRBs with an established SN-association {\footnote {Thirteen GRBs (GRBs~060218A, 071112C, 100316D, 111209A, 111228A, 120714B, 120729A, 130215A, 130831A, 161219B, 171010A, 190114C and 190829A) in the \bat catalog were reported to have a confirmed SN-association \citep{Bufano_2012,Cano_2014,Cano2017,Kann_2019,Klose_2019,Melandri_2022,Hu_2021}. However, GRBs~060218, 071112C, and 171010A were excluded from the sample due to the non-availability of \bat data. GRB~111209A has incomplete data in \bat and was excluded. The final sample has nine GRB-SNe.}} in these embeddings. The SN-associated GRBs occupy the central region of the embedding (central cluster), well separated from the KN-associated GRBs. The only exception is GRB~070714B with a probable KN counterpart, which overlaps with the SN-associated GRBs. Given our sample size and the tentative nature of the detection, it is difficult to investigate on this GRB further. A detailed analysis of the SN-associated GRBs will be the subject of a future paper. 

\cite{Zhu2022} proposed that the KN-associated long GRBs share similar progenitors via analyses of individual GRBs and the features in their light curves. Our analysis using the prompt emission light curves agrees with \citet{Zhu2022} with all KN-associated long GRBs lying in the same cluster. The sub-clustering seen in Fig.~\ref{fig:kilonovae} could be suggestive of something profound. If the KN-associated GRBs in the two clusters are representative of other GRBs in those clusters, then this classification clearly unravels two distinct populations of such GRBs, potentially having different progenitors or formation pathways. 

\subsection{Inference of the physical parameters of the clusters of KN-associated GRBs}
The GRB prompt emission light curves are of paramount importance in understanding the characteristics of the central engine. The \tninty duration describes the time scale over which the central engine is active, and the minimum variability time scale (\mvts) gives an idea about the compactness of the emitting region (\citealt{Sari1997,Golkhou2015,Sonbas_2015}, see also \citealt{Camisasca2023} for a recent work that studied the correlation of \mvts of GRBs with quantities such as peak luminosity and Lorentz factor). We, therefore, explored the effect of these two parameters on the clusters of KN-associated GRBs. The \tninty values are extracted from the \bat catalog. The \mvts values are estimated using the methods described in \cite{Golkhou2015} and \cite{Dimple2022}. 

Fig.~\ref{fig:comparison} shows the PCA-tSNE/UMAP embeddings color-coded with \tninty and \mvts. The two clusters of KN-associated GRBs have a significant difference in their \tninty values. The GRBs lying on the left cluster of the embeddings are of consistently shorter duration than those on the right cluster. On the other hand, both the KN-associated GRB clusters have similar \mvts values.
This suggests that the compactness of the central engines for both populations may be comparable, but they differ in the duration for which the central engine is active. 

In addition, we plot the light curves of KN-associated GRBs belonging to different subclusters identified by \sw{AutoGMM} to examine the similarities between structures of the light curves. Fig. \ref{LCs} shows the light curves in the 15-350 keV band. By simply visualizing these light curves, it is evident that KN-associated GRBs within the same cluster share similar characteristics, such as duration and overall shapes of the light curves, which is further supported by Fig. \ref{fig:comparison}. However, the algorithm uses the light curves in four energy bands and  performs a more thorough job of utilizing the finer structures in the light curves while generating the final maps.

\subsection{Possible progenitors of the KN-associated GRBs}
BNS and/or NS-BH mergers are the most plausible candidates for producing a KN and a GRB ~\citep{Metzger_2010}. There are several possible pathways in these two types of mergers that may lead to very different end states (see Fig. 1 of \citealt{Bartos2013}). These outcomes depend on quantities such as the extent of tidal disruption, masses of the NS and BH that are involved, the maximum mass of the NS that is supported by the equation of state, and spins of the compact binaries involved.

For example, \cite{Gao2022} argued that the mergers of BNSs in which at least one of the NSs has a very high magnetic field, with a magnetic flux of about $10^{29}cm^2\,G$, can lead to halting of the accretion flow well before the innermost stable circular orbit and therefore produce longer duration GRBs. Recent studies by \cite{Lu_2022} also predicted a long-lived accretion disk after the BNS mergers. Similarly, if the BNS merger leads to the formation of a stable hypermassive magnetized NS, the corresponding GRB may be different than those produced from the mergers which directly produces a BH as the remnant (see, for instance, \citealt{Metzger2008,Gompertz2014}). What is even more interesting is that the magnetar in the latter case may be formed not necessarily by a BNS merger but by a double WD merger or accretion-induced collapse of a WD. 

In the case of NS-BH mergers, tidal disruption is favored when BH mass is low, BH spin is prograde and large radii of neutron stars~\citep{Foucart:2018rjc,Foucart2020} (see Fig.~2 of \citealt{Foucart2020}). An efficient tidal disruption produces enough material around the remnant to accrete. Energy injection from the fallback material of the tidally-disrupted NS in a NS-BH merger~\citep{Rees1988, Desai2019} has been suggested as a mechanism that can lead to a longer episode of accretion and hence longer duration GRBs. This may offer an alternative explanation for the GRBs in our right cluster. For instance, \cite{Zhu2022} invoked fallback accretion to the remnant BH to explain long durations of  GRBs 060614, 211211A, and 211227A. 

If these low mass ratio NS-BH mergers are frequent, a relatively nearby population of them may be detectable by LIGO/Virgo detectors in their future observing runs. Though the first LIGO detections of NS-BHs~\citep{LVK-NSBH} did not have any accompanying EM counterparts,
it has been argued by \cite{SarinGRB211211A} that if a LIGO-like detector were operational, GRB~211211A might have been a multi-messenger source if a compact binary merger was its progenitor. It will be interesting to see whether the upcoming fourth observing run (O4) of the LIGO/Virgo interferometers will register the first multi-messenger NS-BH merger. When we have a handful of multimessenger compact binary mergers with associated GRBs/KNe, a more robust test of this classification hypothesis can be carried out. 

As the next-generation GW detectors, such as Voyager~\citep{2020CQGra..37p5003A}, Cosmic Explorer~\citep{evans2021horizon}, Einstein Telescope etc.~\citep{2012CQGra..29l4013S,kalogera2021generation}, become operational, the large population of BNS and NS-BH mergers they will detect should help in uncovering the bigger landscape of the GRB population and answer several open problems related to the progenitor physics. More sensitive optical instruments, such as Vera Rubin Observatory, may also enhance our horizons for KN detections much beyond what we have today~\citep{2023arXiv230108763G}. 

In summary, in this {\it Letter}, we show there exist two distinct classes of KN-associated GRBs. Only future multimessenger observations can reveal whether they are two classes of BNS or NS-BH mergers or something else. There are several possible future directions of this work. Analysis of the {\it Fermi} catalog of GRBs, similar to \cite{Steinhardt2023} but tracking the presence of GRBs associated with SNe and KNe to gain insights and the inclusion of afterglow information into the current analysis are two of them. As machine-learning-based methods continue to evolve and improve, they will likely play an increasingly important role in understanding the progenitors of GRBs.

\section*{acknowledgments}
We thank the referee for the critical reading of the manuscript and the comments that have improved the presentation of the manuscript. We thank L. Resmi for the delightful discussions and very useful comments on the manuscript. KGA thanks Ish Gupta, Arnab Dhani, and Rahul Kashyap for useful discussions. K.G.A. acknowledges the Swarnajayanti grant DST/SJF/PSA-01/2017-18 of the Department of Science and Technology and SERB, India, and support from Infosys Foundation.

\bibliography{main}{}
\bibliographystyle{aasjournal}

\appendix
\counterwithin{figure}{section}
\section{Similarity between PCA-tSNE/UMAP}
\label{comparison_flameplot}
Although both algorithms, PCA-tSNE/UMAP, excel in dimensionality reduction, they are sensitive to hyperparameters, which can significantly affect the embeddings produced. To ensure optimal performance, we carefully tuned the hyperparameters for each algorithm separately. However, it is crucial to compare the results of both algorithms to ensure their consistency and reliability. 
We quantify the similarities between the two maps using the python module \sw{flameplot} developed by \cite{Taskesen_flameplot_is_a_2020}. Flameplot works by quantifying the local similarities across two maps or embeddings.
For the comparison of two embeddings, X and Y, the algorithm compares the kX and kY nearest neighbors of each embedding. The algorithm first computes the Euclidean distance between samples in each embedding and ranks the sample using Euclidean distance with the smallest distance at the top. Next, it compares the ranks of embedding X to the ranks of embedding Y for the nearest neighbors of kX and kY. Finally, it quantifies the score for the overlap between ranks. The score is a measure of random similarity between cards. If the two maps completely overlap, the score is 1. 

\begin{figure*}[!h]
\centering
    \includegraphics[width=0.5\columnwidth]{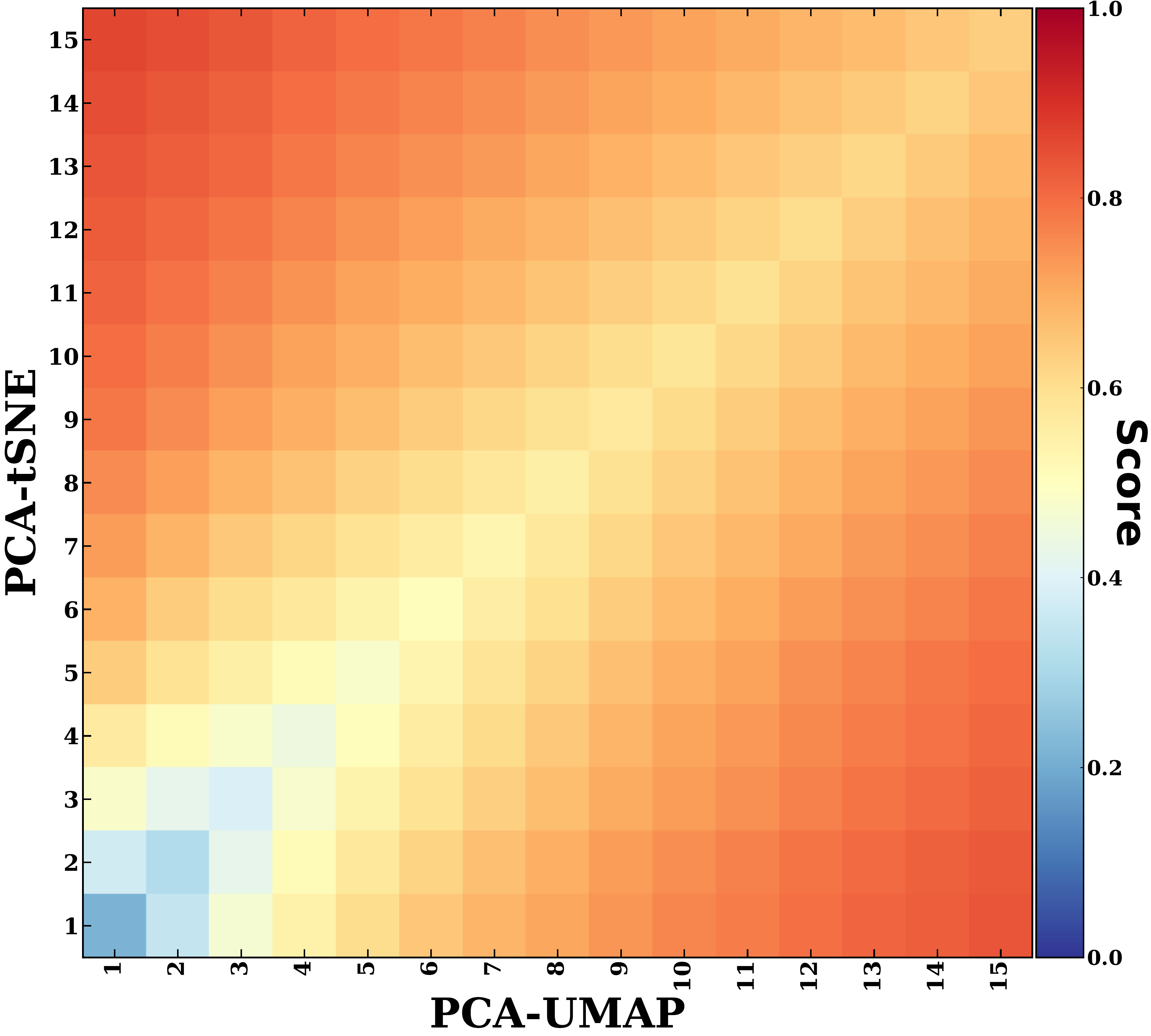}
    \caption{The heatmap illustrates the similarity score between PCA-tSNE and PCA-UMAP embeddings, which is close to 1 for a high value of nearest neighbors. Despite a decrease in score with fewer neighbors, it still stays above 0.5, suggesting that the two embeddings, initialized by PCA, are similar. 
        }
    \label{flameplot}
\end{figure*}

The heatmap in Fig.~\ref{flameplot} shows the similarity score (a measure of random similarity, where a higher score means more similar embeddings) for the 15 nearest neighbors. Despite a decrease in score with fewer neighbors, it remains above 0.5, demonstrating that PCA-initialized tSNE and UMAP embeddings are similar.

\section{Light curves of KN-associated GRBs}
\begin{figure*}
\centering
    \includegraphics[width=0.32\columnwidth]{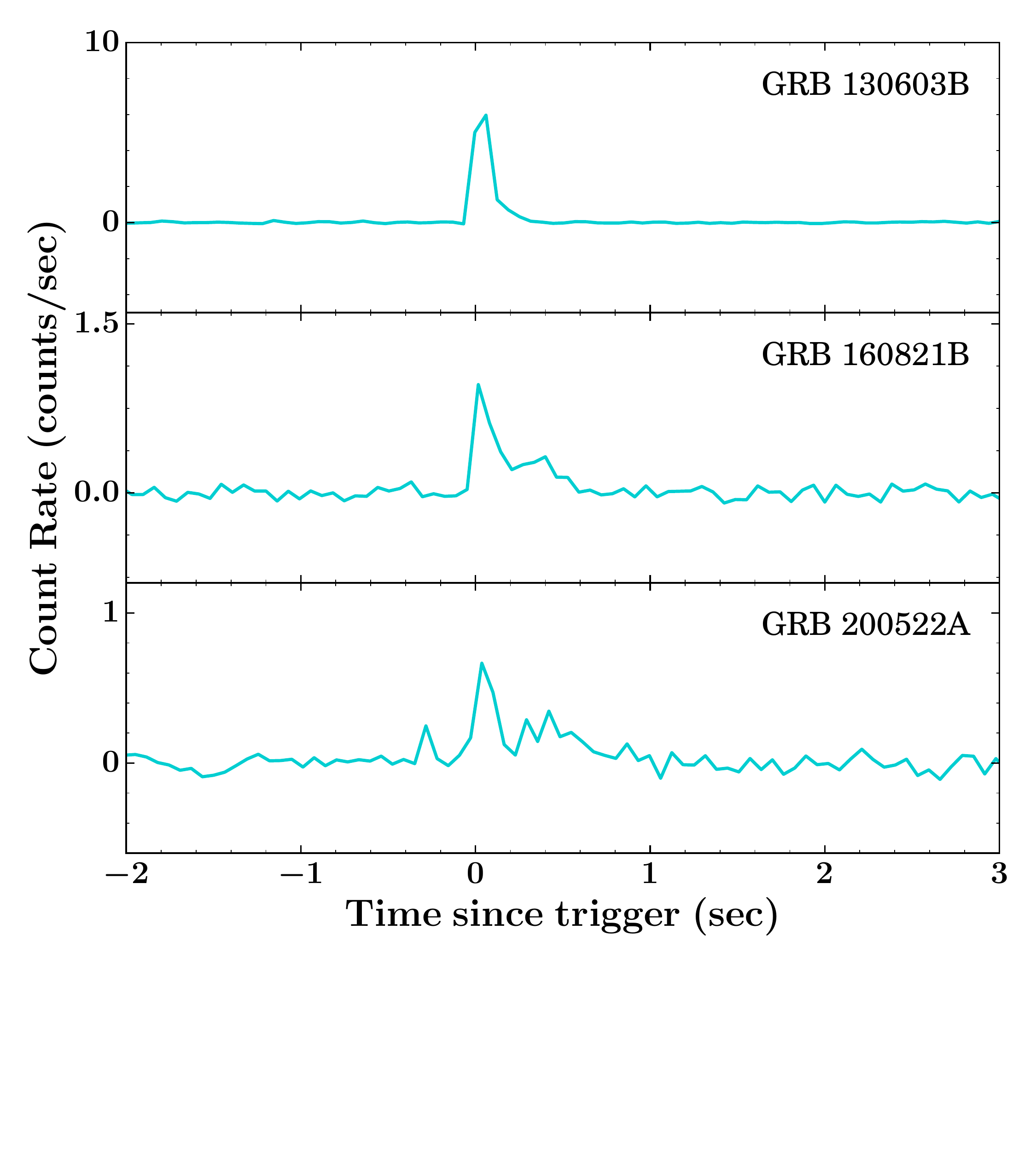}
    \includegraphics[width=0.32\columnwidth]{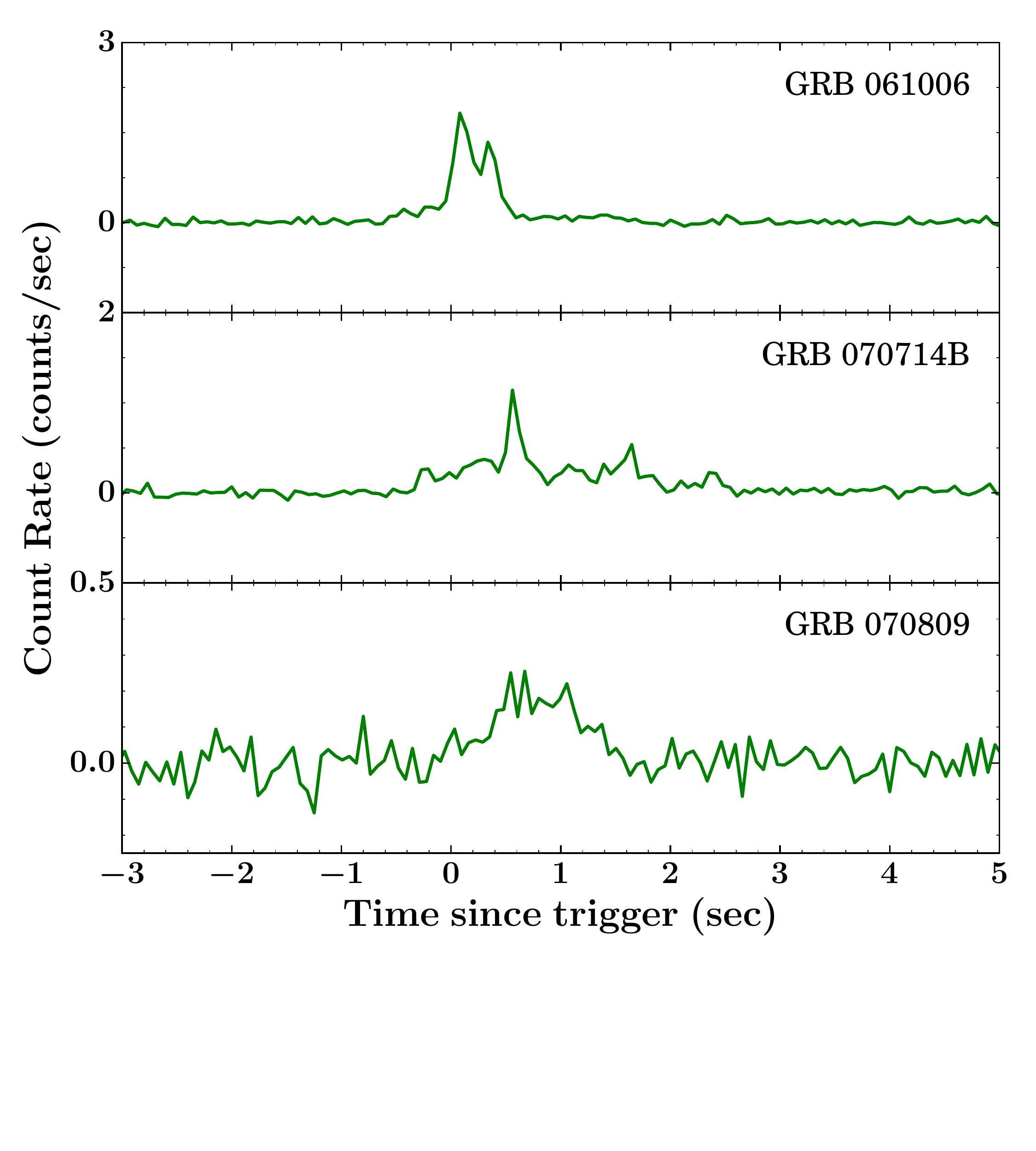}
    \includegraphics[width=0.32\columnwidth]{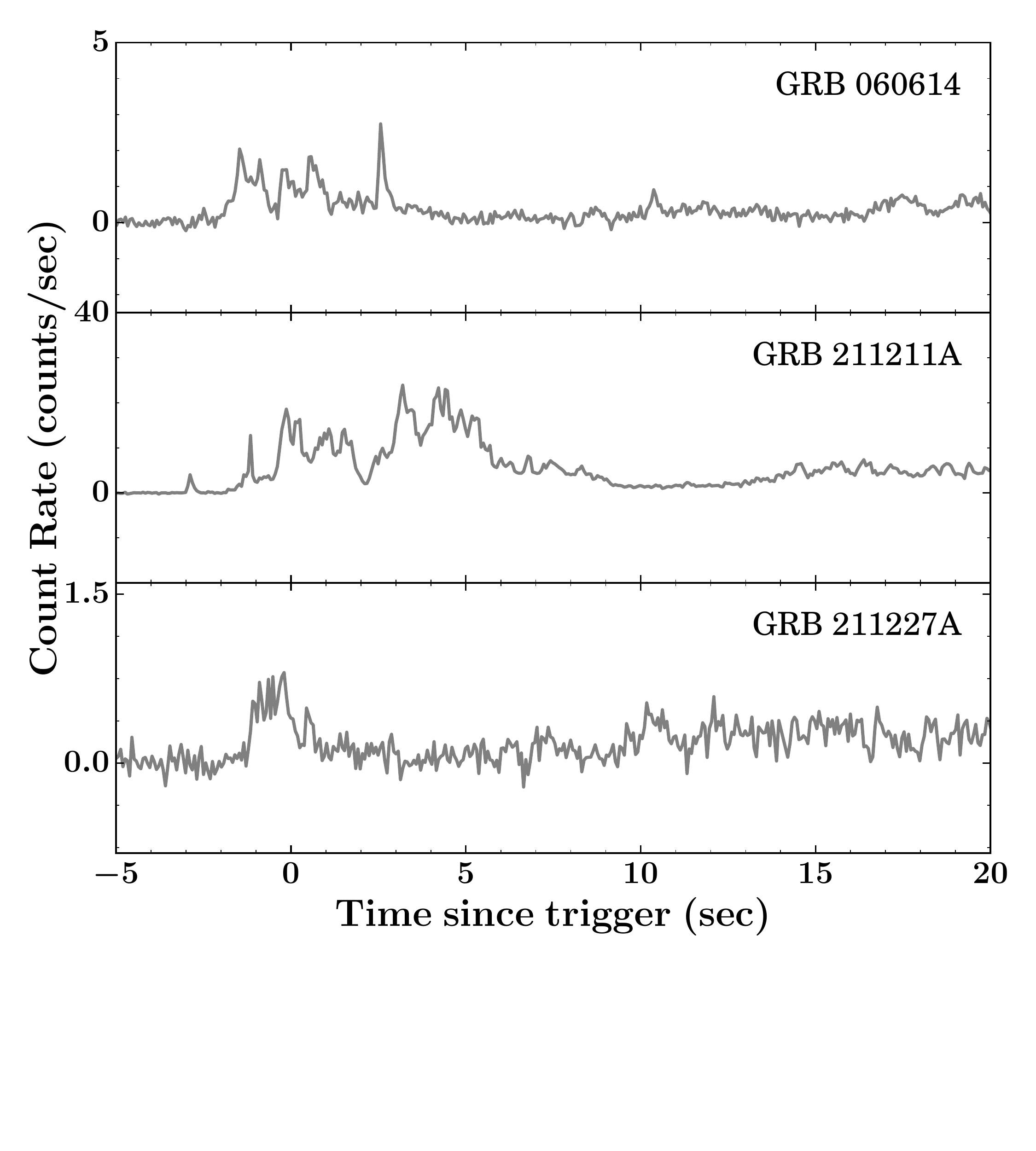}
    \caption{The light curves of KN-associated GRBs belonging to various subclusters as identified by AutoGMM.}
    \label{LCs}
\end{figure*}

The algorithm processes four light curves and captures collective images of these curves across different energy bands. However, to avoid the complexity, we display only the light curves  of the KNe groups in the 15-350 keV band. Fig. \ref{LCs} shows the light curves of the kilonovae belonging to various subgroups ([GRB 130603B, 160821B, and 200522A, upper left group, turquoise colored], [GRB 060614, 211211A and 211227A, upper right  group, grey colored], [GRB 061006, 070714B and 070809, group colored as green] in Fig. \ref{fig:kilonovae}).

\end{document}